\documentclass[aps,prl,preprint,groupedaddress]{revtex4-1}

\usepackage{amsmath}
\usepackage{amssymb}
\usepackage{graphicx}
\usepackage{mathrsfs}
\usepackage{lineno}
\usepackage{bbm}
\usepackage{bbold}
\usepackage{float}

\begin{document}

\title{Exact derivation and practical application of a hybrid stochastic simulation algorithm for large gene regulatory networks}

\author{Jaroslav Albert\\
jaroslavalbert81@gmail.com}
\begin{abstract}
We present a highly efficient and accurate hybrid stochastic simulation algorithm (HSSA) for the purpose of simulating a subset of biochemical reactions of large gene regulatory networks (GRN). 
The algorithm relies on the separability of a GRN into two groups of reactions, A and B, such that the reactions in A can be simulated via a stochastic simulation algorithm (SSA), while those
in group B can yield to a deterministic description via ordinary differential equations. First, we derive exact expressions needed to sample the next reaction time and reaction type,
and then give two examples of how a GRN can be partitioned. Although the methods presented here can be applied to a variety of different stochastic systems within GRN, we focus on simulating
mRNAs in particular. To demonstrate the accuracy and efficiency of this algorithm, we apply it to a three-gene oscillator, first in
one cell, and then in an array of cells (up to 64 cells) interacting via molecular diffusion, and compare its performance to the Gillespie algorithm (GA). 
Depending on the particular numerical values of the system parameters, and the partitioning itself, we show that our algorithm is between 11 and 445 times faster than the GA.
\end{abstract}

\maketitle

\section{introduction}

The business of modeling chemical reaction networks generally falls into two camps: 1) chemical master equation (CME); and 2) stochastic simulation algorithms (SSA).
Historically, research in the former has centered on solving the CME using analytic methods, both exact and approximate \cite{Jahnke, Pendar, Swain, Walczak, Bokes, Popovic}, or by developing efficient numerical techniques \cite{Mugler, Wolf, Albert, Albert2, Albert3, Bokes2, Veerman, Munsky, Gupta}.
The latter, popularly known and the Gillespie algorithm (GA) \cite{Gillespie}, has emerged as an alternative to the CME and is generally more preferred for 
modeling systems with many species and/or reaction channels. Although the GA is simple in construction and guarantees
a solution, it tends to be computationally expensive. Hence, research on faster versions of the GA has been ongoing \cite{Gibson, Gillespie2, Cao, Cao2, Cao3}. Although the choice between the CME and an SSA depends on  
several factors, such as the complexity of the reaction network, efficiency of each approach etc., the general rule is this: use the CME if it can be solved either analytically or numerically; otherwise employ an SSA. In gene regulatory networks (GRN), which will be the focus of this paper, most of the systems of interest require the use of an SSA.

In recent years, a new camp has emerged in which the CME and the GA are combined into a hybrid \cite{Haseltine,Burrage, Salis, Jahnke2, Albert4, Albert5, Zechner, Duso, Kurasov}. The purpose of hybrid models is to avoid stochastically simulating every reaction and 
instead to use ordinary differential equations (ODE), e. g. the CME, to describe a subset of reactions, which can lead to a significant increase in computational speed. If the ODE for such a subset can be solved efficiently (preferably analytically), then the numerical cost would be incurred only by the remaining reactions. The first hybrid algorithm was proposed by Haseltine and  Rawlings \cite{Haseltine}, and later improved upon by Burrage {\it et. al.} \cite{Burrage} and Salis and Kaznessis \cite{Salis}.
The principle behind both works is the segregation of reactions into fast and slow reactions, wherein the fast reactions are simulated by the Lengevin equation, while the slow reactions are handled by an SSA. The applicability of this algorithm, of course, is limited
to systems that contain both fast and slow reactions. Jahnke and Altıntan \cite{Jahnke2} developed a hybrid method that splits the system into two (or more)
subsets of reactions in such a way that the CME for one subset can be solver analytically, while the reactions in the other subset are simulated in a $\tau$-leaping fashion (see reference \cite{Gillespie2}). Later on, other hybrid models, ones that could be implemented for any set of reactions, fast or otherwise, were developed
\cite{Albert4, Albert5, Zechner, Duso, Kurasov}. In these later models, the CME is used to describe a subset of reactions, while a modified SSA is applied to the remaining set. The main obstacle to this approach is to work out
the probability distributions for the next reaction time and the reaction index of the latter subset, as they both depend on the state history of the species described by the CME. One way to solve this issue
is to treat the first subset as deterministic, thereby collapsing the system's many possible histories into one, as was done in \cite{Kurasov}. However, this approach restricts the ways a system can be partitioned into those that lend themselves to such approximations. Duso and Zechner \cite{Duso}
developed a hybrid model that coupled statistical moments of the chemical species in the first subset to the probabilities of the reaction time and reaction type of the latter subset.

In this paper, we build on our earlier work \cite{Albert4, Albert5} in which we derived the probability
for the next reaction time and the probabilities for the reaction types for the letter reaction subset. However, the method was demonstrated only on systems without 
any reactions in which two molecules combine into another molecular species, e.g. dimerization. In the present work, we demonstrate how any reaction
network can be partitioned into a subset of reactions describable by a CME and a subset that is simulated with an SSA by deriving exact: probability distribution for the next
reaction time; probability for the reaction type; and the conditional joint probability for all the species in the first subset, given the sampled next reaction time and reaction type.
Although the formalism of our approach is exact and works for any reaction network, the application of it varies in difficulty depending on the manner in which a network
is partitioned. Therefore, to demonstrate our method in practice, we show two examples of how a system of three interacting genes can be partitioned in order to render
all calculations tractable. The dynamics of the system we chose is oscillatory, which is suitable for testing the effects of systematic errors, as these types of error would manifest
through phase shifts.
Our results show excellent accuracy for all parameter sets and an increase in efficiency by factors ranging from 11 to 445.


\section{Master equation and the Gillespie algorithm}

Given a set of molecular copy numbers ${\bf x} = \{x_1,...,x_V\}$, and a set of reactions
\begin{equation}\label{reaction_events}
\sum_{i=1}^V\alpha^i_{\mu}x_i\xrightarrow{\makebox[1cm]{$a_{\mu}({\bf x},t)$}}\sum_{i=1}^V\beta^i_{\mu}x_i\,\,\,\,\,\mu=1,2,...,J
\end{equation}
where the integers $\alpha^i_k$ and $\beta^i_k$ are the stoichiometric coefficients and $a_{\mu}({\bf x},t)$ are the reaction propensities,
the joint probability distribution, $P({\bf x},t)$, for ${\bf x}$ is a solution of the chemical master equation (CME)
\begin{equation}\label{ME}
\frac{dP({\bf x},t)}{dt}=\sum_{\mu=1}^Ja_{\mu}({\bf x}-{\bf f}_{\mu},t)P({\bf x}-{\bf f}_{\mu},t)-P({\bf x},t)\sum_{\mu=1}^Ja_{\mu}({\bf x},t).
\end{equation}
The state-change vector ${\bf f}_{\mu}$ specifies the change in ${\bf x}$ due to the ${\mu}^{\text{th}}$ reaction: 
${\bf x}\rightarrow {\bf x}+{\bf f}_{\mu}$.
Except for a very specific class of systems, Eq. (\ref{ME}) cannot be solved exactly. Attempts to solve it
approximately, using either analytic or numerical methods, are usually thwarted by the {\it curse of dimensionality}, 
which tends to cast itself on many-variable systems. 

An alternative approach is to simulate the reaction events in Eq. (\ref{reaction_events}) by sampling the time between reactions, $\tau$, and
the reaction index $\mu=1,...,J$ from a joint probability distribution $P(\mu;\tau)$. The time is then advanced by $\tau$ and the system variables are
updated based on which reaction took place.
This procedure is called the Gillespie algorithm (GA) and it can be derived as follows:

The probability that no reaction occurs within an infinitesimal time interval $dt$ is given by
\begin{equation}
P(dt)=(1-R({\bf x},0)dt),
\end{equation}
where
\begin{equation}
R({\bf x},t)=\sum_{\mu}a_{\mu}({\bf x},t).
\end{equation}
The probability that no reaction occurs within a finite time $t=dtN$ is simply
\begin{equation}\label{Prob_tau}
P(t)=\prod_{n=1}^N(1-R({\bf x},t_n)dt)=\text{exp}\left[-\int_0^tdt'R({\bf x},t')dt\right],
\end{equation}
where $t_n=ndt$.
The probability that no reaction occurs up to $t$ and that reaction $\mu$
occurs between $t$ and $t+dt$ is given by
\begin{equation}
P({\mu};t)=P(t)a_{\mu}({\bf x},t)dt.
\end{equation}
Since $t$ can be sampled independently of $\mu$, via Eq. (\ref{Prob_tau}),
$\mu$ must be sampled from a conditional probability $P(\mu|\tau)$ that $\mu$ occurs
provided $t$ has been observed to be some time $\tau$. According to the Bayes relation, $P(\mu|\tau)$ can be expressed as
\begin{equation}
P(\mu|\tau)=\frac{P({\mu};\tau)}{\sum_{\nu}P({\nu};\tau)}=\frac{a_{\mu}({\bf x},\tau)}{R({\bf x},\tau)}.
\end{equation}
For systems in which the reaction propensities are time-independent, the two probabilities reduce to
the well known expressions:
\begin{eqnarray}\label{GA_prob}
&&P(\tau)=e^{-R({\bf x})\tau}\,\,\,\,\,\,\,\,\,\,\,\,\,\,\,\,\,\,\text{probability to observe $t=\tau$}\label{Pt}\\
&&P(\mu|\tau)=\frac{a_{\mu}({\bf x})}{R({\bf x})}\,\,\,\,\,\,\,\,\,\,\,\,\,\,\,\,\text{probability to observe reaction $\mu$, given $\tau$}\label{Pt}\label{Pmu}
\end{eqnarray}
With these simple relations, the GA can be implemented as follows:
\vspace{0.5 cm}
\newline
\hspace*{0.5 cm}1. At $t=0$ choose an initial state ${\bf x}$ and compute the propensities $a_{\mu}({\bf x})$.
\newline
\hspace*{0.5 cm}2. Select two random numbers $\xi_1$ and $\xi_2$.
\newline
\hspace*{0.5 cm}3. Compute $\tau$ by solving $P(\tau)=\xi_1$.
\newline
\hspace*{0.5 cm}4. Find the smallest integer $\mu$ that satisfies
\begin{equation}
\hspace*{0.5 cm}\sum_{\nu=1}^{\mu}P(\nu|\tau)>\xi_2.
\end{equation}
\newline
\hspace*{0.5 cm}5. Update the variables by letting ${\bf x}\rightarrow{\bf x}+{\bf f}_{\mu}$ and set $t$
to $t+\tau$.
\newline
\hspace*{0.5 cm}6. Return to step 1. 
\vspace{0.5 cm}

A solution to Eq. (\ref{ME}) can be obtained by running the GA enough times to generate a
statistically significant ensemble. While this process guarantees a solution, it can be very time consuming.

\section{A hybrid approach}

Let us consider a reaction network that can be partitioned into two sets of reactions, $A=\{a_1({\bf x}),...,a_{\kappa}({\bf x})\}$ 
and $B=\{a_{\kappa+1}({\bf x}),...,a_J({\bf x})\}$, such that a subset of variables ${\bf y}=\{y_1=x_1,...,y_M=x_M\}$ is affected
only by reactions in $A$, while the remaining set ${\bf z}=\{z_1=x_{M+1},...,z_{V-M}=x_V\}$ can be affected by reactions in both $A$ and $B$.
We define ${\bf g}_{\mu}$ as the state-change vector that affects ${\bf y}$ only, and ${\bf h}_{\mu}$ as the state-change vector that affects exclusively ${\bf z}$.
 
Let us now imagine that we know the exact stochastic evolution of the variable set ${\bf z}$
during some time $t$; in other words, we know the path of each variable in ${\bf z}$ in the $t-z$ plane.
We could then ask: what are the probabilities that 1) no reaction in $A$ occurs for
during $t$; and 2) reaction $\mu$ occurs between $t$ and $t+dt$? The former can be constructed the same way as in Eq. (\ref{Prob_tau}):
\begin{equation}\label{Ptz}
P(t|{\bf Z})=\prod_{n=1}^N[1-R_A({\bf z}(t_n),{\bf y})dt]=\text{exp}\left[-\int_0^{t}R_A({\bf z}(t'),{\bf y})dt'\right],
\end{equation}
where
\begin{equation}
R_A({\bf z}(t),{\bf y})=\sum_{\nu=1}^{\kappa}a_{\nu}({\bf z}(t),{\bf y})
\end{equation}
and ${\bf Z}$ is the set of paths taken by the variables in ${\bf z}$, i.e. 
\begin{equation}
{\bf Z}=\underset{\underset{\text{\normalsize{path of}}\,\,\scalebox{1}{$z_1$}\,\,\,\,\,\,\,\,\,\,\,\,\,\,\,\,\,\,\,\,\,\,\,\,\,\,\,\,\,\text{\normalsize{path of}}\,\,\scalebox{1}{$z_2$}\,\,\,\,\,\,\,\,\,\,\,\,\,\,\,\text{\normalsize{...}}\,\,\,\,\,\,\,\,\,\,\,\,\,\,\,\,\,\,\,\,\,\,\,\,\,\,\,\,\,\,\,\,\,\,\,\,\,\,\,\,\,\,\,\,\,\,\,\,\,\,\,\,\,\,\,\,\,\,\,\,\,\,\,\,\,\,\,\,\,\,\,\,\,\,\,\,\,\,\,\,\,\,\,\,\,\,\,\,\,\,\,\,\,\,\,\,\,\,\,\,}{}}{\{\underbrace{z_1(t_0),...,z_1(t_N)};
\underbrace{z_2(t_0),...,z_2(t_N)};...;z_{{V-M}}(t_0),...,x_{{V-M}}(t_N)\}}
\end{equation}
Since we do not actually know the paths, i. e. the values of all the entries in ${\bf Z}$, we must multiply
Eq. (\ref{Ptz}) by the probability for ${\bf Z}$, ${\cal P}({\bf Z})$, and sum over all the variables $z_i(t_n)$, except for the endpoints of each path, ${\bf z}(0)={\bf z}_0$ and ${\bf z}(t)={\bf z}_t$, 
which we keep fixed:
\begin{equation}\label{Qztz0}
Q({\bf z}_t,t|{\bf z}_0)=\sum_{{\bf Z}\neq{\bf z}_0,{\bf z}_t}{\cal P}({\bf Z})P(\tau|{\bf Z}).
\end{equation}
If the initial set ${\bf z}_0$ is known only with some probability $P_{\text{in}}({\bf z}_0)$, we must multiply Eq. (\ref{Qztz0}) by $P_{\text{in}}({\bf z}_0)$ and sum over ${\bf z}_0$:
\begin{equation}\label{Qzt}
Q({\bf z}_t,t)=\sum_{{\bf z}_0}P_{\text{in}}({\bf z}_0)Q({\bf z}_t,t|{\bf z}_0).
\end{equation}
We will refer to Eq. (\ref{Qzt}) as the ${\it Q}$-distribution; it is the joined probability that no reaction occurs until $t$ and that the variables ${\bf z}$ will have the values
${\bf z}_t$ at $t$. Finally, the probability that no reaction occurs until $t$ is given by
\begin{equation}\label{Qt}
Q(t)=\sum_{{\bf z}_t}Q({\bf z}_t,t).
\end{equation}

Next, we would like to compute the conditional probability, $Q(\mu|t)$, that reaction $\mu$ occurs between $t$ and $t+dt$, provided no reaction occured until $t$.
This can be obtained from the joined probability, $Q({\bf z}_t,\mu,t)$, that no reaction occurs until 
$t$, that ${\bf z}={\bf z}_t$ at $t$, and that reaction $\mu$ occurs between $t$ and $t+dt$, which is simply this:
\begin{equation}
Q({\bf z}_t,\mu,t)=Q({\bf z}_t,t)a_{\mu}({\bf z}_t,{\bf y})d\tau.
\end{equation}
So, the conditional probability that reaction $\mu$ occurs, provided no reaction occured up to $t$, reads
\begin{equation}\label{Qmut}
Q(\mu|t)=\frac{\sum_{{\bf z}_t}Q({\bf z}_t,\mu,t)}{\sum_{{\bf z}_t,\mu}Q({\bf z}_t,\mu,t)}.
\end{equation}
To update the initial probability, $P_{\text{in}}({\bf z}_0)$, after $t$ and $\mu$ are sampled, we must compute the conditional probability
to observe ${\bf z}_t$, provided no reaction occured until $t$ and reaction $\mu$ occured at $t$, which is given by
\begin{equation}\label{Qzmut}
Q({\bf z}_t|,\mu,t)=\frac{Q({\bf z}_t,\mu,t)}{\sum_{{\bf z}_t}Q({\bf z}_t,\mu,t)}
=\frac{Q({\bf z}_t,t)a_{\mu}({\bf z}_t)}{\sum_{{\bf z}_t}Q({\bf z}_t,t)a_{\mu}({\bf z}_t)},
\end{equation}
and set ${\bf z}_t={\bf z}_0$. Note that this ``update" only considers how $P_{\text{in}}({\bf z}_0)$ changes given new information,
namely that reaction $\mu$ has occurred at $t=\tau$. However, we must also take into account how the reaction itself affects $P_{\text{in}}({\bf z}_0)$.
For example, if a transcription factor (TF) bonded to a promoter, there would be one less TF available to bind to promoters. The probability of there being
$n$ number of available TFs would be shifted to the left along the $n$-axis. This shift is simple when the distribution is far enough from zero, i. e. when $P(n=0)\ll 1$:
$P(n)\rightarrow P(n+1)$. For a general case, one must shift $P(n)$ to the left by one, but add $P(n=1)$ to $P(n=0)$: $P(n=0)\rightarrow P(n=0)+P(n=1)$ and  
$P(n=i)\rightarrow P(n=i+1)$ for $i>0$. On the other hand, when a TF dissociates form a promoter, thereby adding a TF to the system, the shift is always $P(n)\rightarrow P(n-1)$.
For breavity, we define a shifting operator ${\hat S}_{\mu}$, whose action will update appropriately $P_{\text{in}}({\bf z}_0)$ as a function of
$\mu$.

The last piece of the puzzle is to figure out how to compute Eq. (\ref{Qzt}). Fortunately, this has already been done in a previous
work of ours \cite{Albert4}, in which we showed that
Eq. (\ref{Qzt}) is the solution of this equation:
\begin{equation}\label{ModME}
\frac{dQ({\bf z},t)}{dt}=\sum_{\nu={\kappa+1}}^Ja_{\nu}({\bf z}-{\bf h}_{\nu},t)Q({\bf z}-{\bf h}_{\nu},t)-Q({\bf z},t)\sum_{\nu={\kappa+1}}^Ja_{\nu}({\bf z},t)
-R_A({\bf z},{\bf y})Q({\bf z},t)
\end{equation}
with the initial conditions $Q({\bf z},0)=P_{\text{in}}({\bf z})$.
Note that if we set $R_A$ to zero, Eq. (\ref{ModME}) would reduce to the CME for $B$. Hence, we will refer to Eq. (\ref{ModME})
as {\it modified chemical master equation} (MCME).

We now have all the ingredients to implement the hybrid stochastic simulation algorithm (HSSA).
The steps are as follows:
\vspace{0.5 cm}
\newline
\hspace*{0.5 cm}1. At $t=0$ choose an initial state ${\bf y}$ and initial probability $P_{\text{in}}({\bf z})$.
\newline
\hspace*{0.5 cm}2. Select two random numbers $\xi_1$ and $\xi_2$.
\newline
\hspace*{0.5 cm}3. Solve Eq. (\ref{ModME}) for $Q({\bf z},t)$, and compute $Q(t)$, $Q(\nu|t)$ and $Q({\bf z}_t|,\mu,\tau)$ from Eqs. (\ref{Qt}), 
\hspace*{1 cm}(\ref{Qmut}) and (\ref{Qzmut}),
respectively.
\newline
\hspace*{0.5 cm}4. Compute $\tau$ by solving $Q(\tau)=\xi_1$.
\newline
\hspace*{0.5 cm}5. Find the smallest integer $\mu$ that satisfies
\begin{equation}
\hspace*{0.5 cm}\sum_{\nu=1}^{\mu}Q(\nu|\tau)>\xi_2.
\end{equation}
\newline
\hspace*{0.5 cm}6. Update the variables ${\bf y}$ by letting ${\bf y}\rightarrow{\bf y}+{\bf g}_{\mu}$ and set $t$
to $t+\tau$.
\newline
\hspace*{0.5 cm}7. Update the initial probability: $P_{\text{in}}({\bf z})={\hat S}_{\mu}Q({\bf z}|,\mu,\tau)$.
\newline
\hspace*{0.5 cm}8. Return to step 1. 
\vspace{0.5 cm}
\newline

The above steps give a recipe on how to implement an HSSA. However, step 3 relies on one important assumption, namely, that Eq. (\ref{ModME}) is tractable.
This is where the manner in which a reaction network is partitioned becomes important for practical reasons. If the number and nature of reactions in $B$ are
such that Eq. (\ref{ModME}) can be solved either analytically or numerically, then the HSSA can be implemented in practice. Furthermore, since the purpose
of deriving the HSSA is to outperform conventional SSAs, such as the GA, step 3 must not only be feasible, but needs to be done efficiently. These considerations
will essentially dictate how a reaction networks is partitioned. In the next section, we give two examples of how a reaction system can be partitioned. The last section will be reserved
for discussing the performances of said examples.

\section{Partitioning of the system: practical applications}

To make our examples concrete, we will focus exclusively on gene regulatory networks (GRN). 

The system we want to partition, shown in Fig. 1, is made up of the following variables:
\begin{eqnarray}\label{reactions}
&&w^k_j\,\,\,\,\,\,\,\,\,\,\,\,\,\,\,\,\,\,\text{promoter of gene $k$ in state $j$}\,\,\,\,\,\,\,\,\,\,\,\,\,\,\,\,\,\,\,\,\,\,\,\,\,\,\,\,\,\,\,\,\,\,\,\,\,\,\,\,\,\,\,\,\,\,\,\,\,\,\,\,\,\text{range: $0$ or $1$}\nonumber\\
&&m^k\,\,\,\,\,\,\,\,\,\,\,\,\,\,\,\,\,\text{mRNA copy number of gene $k$}\,\,\,\,\,\,\,\,\,\,\,\,\,\,\,\,\,\,\,\,\,\,\,\,\,\,\,\,\,\,\,\,\,\,\,\,\,\,\,\,\,\,\,\,\,\,\,\,\,\,\text{range: $0$ to $\infty$}\nonumber\\
&&n^k\,\,\,\,\,\,\,\,\,\,\,\,\,\,\,\,\,\,\,\text{protein copy number of gene $k$}\,\,\,\,\,\,\,\,\,\,\,\,\,\,\,\,\,\,\,\,\,\,\,\,\,\,\,\,\,\,\,\,\,\,\,\,\,\,\,\,\,\,\,\,\,\,\,\,\,\text{range: $0$ to $\infty$}\nonumber\\
&&s^{k}\,\,\,\,\,\,\,\,\,\,\,\,\,\,\,\,\,\,\,\text{dimer copy number composed of 2$n^k$}\,\,\,\,\,\,\,\,\,\,\,\,\,\,\,\,\,\,\,\,\,\,\,\,\,\,\,\,\,\,\,\,\,\,\text{range: $0$ to $\infty$};\nonumber\\
\end{eqnarray}
and its evolution is driven by these reactions:
\begin{eqnarray}\label{reactions}
&&1.\,\,\,w^k_j+s^{q}\longrightarrow w^k_{j'}\,\,\,\,\,\,\,\,\,\,\,\,\,\,\,\,\,\,\,a_k^{jj'}w^k_js^{q}\nonumber\\
&&2.\,\,\,w^k_{j'}\longrightarrow w^k_j+s^{q}\,\,\,\,\,\,\,\,\,\,\,\,\,\,\,\,\,\,\,b_k^{jj'}w^k_{j'}\nonumber\\
&&3.\,\,\,\emptyset\longrightarrow m^k\,\,\,\,\,\,\,\,\,\,\,\,\,\,\,\,\,\,\,\,\,\,\,\,\,\,\,\,\,\,\,\,\,\,\,r^1_kw^k_1\nonumber\\
&&4.\,\,\,m^k\longrightarrow \emptyset\,\,\,\,\,\,\,\,\,\,\,\,\,\,\,\,\,\,\,\,\,\,\,\,\,\,\,\,\,\,\,\,\,\,\,d^m_km^k\nonumber\\
&&5.\,\,\,m^k\longrightarrow m^k+n^k\,\,\,\,\,\,\,\,\,\,\,\,\,\,\,\,\,K^m_km^k\nonumber\\
&&6.\,\,\,n^k\longrightarrow \emptyset\,\,\,\,\,\,\,\,\,\,\,\,\,\,\,\,\,\,\,\,\,\,\,\,\,\,\,\,\,\,\,\,\,\,\,\,\,d^n_kn^k\nonumber\\
&&7.\,\,\,2n^k\longrightarrow s^{k}\,\,\,\,\,\,\,\,\,\,\,\,\,\,\,\,\,\,\,\,\,\,\,\,\,\,\,\,\,\,\,\,c^{+}_{k}n^k(n^k-1)/2\nonumber\\
&&8.\,\,\,s^{k}\longrightarrow 2n^k\,\,\,\,\,\,\,\,\,\,\,\,\,\,\,\,\,\,\,\,\,\,\,\,\,\,\,\,\,\,\,\,c^{-}_{k}s^{k}\nonumber\\
&&9.\,\,\,s^{k}\longrightarrow \emptyset\,\,\,\,\,\,\,\,\,\,\,\,\,\,\,\,\,\,\,\,\,\,\,\,\,\,\,\,\,\,\,\,\,\,\,\,\,\,d^s_{k}s^{k}.
\end{eqnarray}
Reaction 1 takes a promoter of gene $k$ from the state labeled by $j$ to the state labeled by $j'$; while reaction 2 is the reverse of reaction 1 
(in general, the state label $j'$ in $a_k^{jj'}$ and $b_k^{jj'}$ depends on the index $q$). Reactions 3 and 4 synthesize and degrade $m^k$, respectively.
Reactions 5 and 6 synthesize and degrade $n^k$, respectively.
Reactions 7 and 8 bind two copies of $n^k$ to form one copy of $s^{k}$ and dissociate $s^{k}$ into two copies of $n^k$, respectively. Finally, reaction 9 degrades $s^{k}$.
The list of reactions (\ref{reactions}) is very standard in systems biology, but is not exhaustive. We could have added the formation of (homo/hetero) trimers, quadrimers and higher oligamers,
post-transcription and post-translation processes, transport across the nuclear membrane and others. For simplicity however, we will stick to the reactions in (\ref{reactions}).

In the following examples, we will focus on obtaining the stochastic behavior of the variable set ${\bf y}$ only; in the process, the information about the set ${\bf z}$
will be relegated to its average: $\langle{\bf z}\rangle$.

\subsection{Example 1}

\begin{figure}
\centering
\includegraphics[trim=0 0 0 1.0cm, height=0.45\textheight]{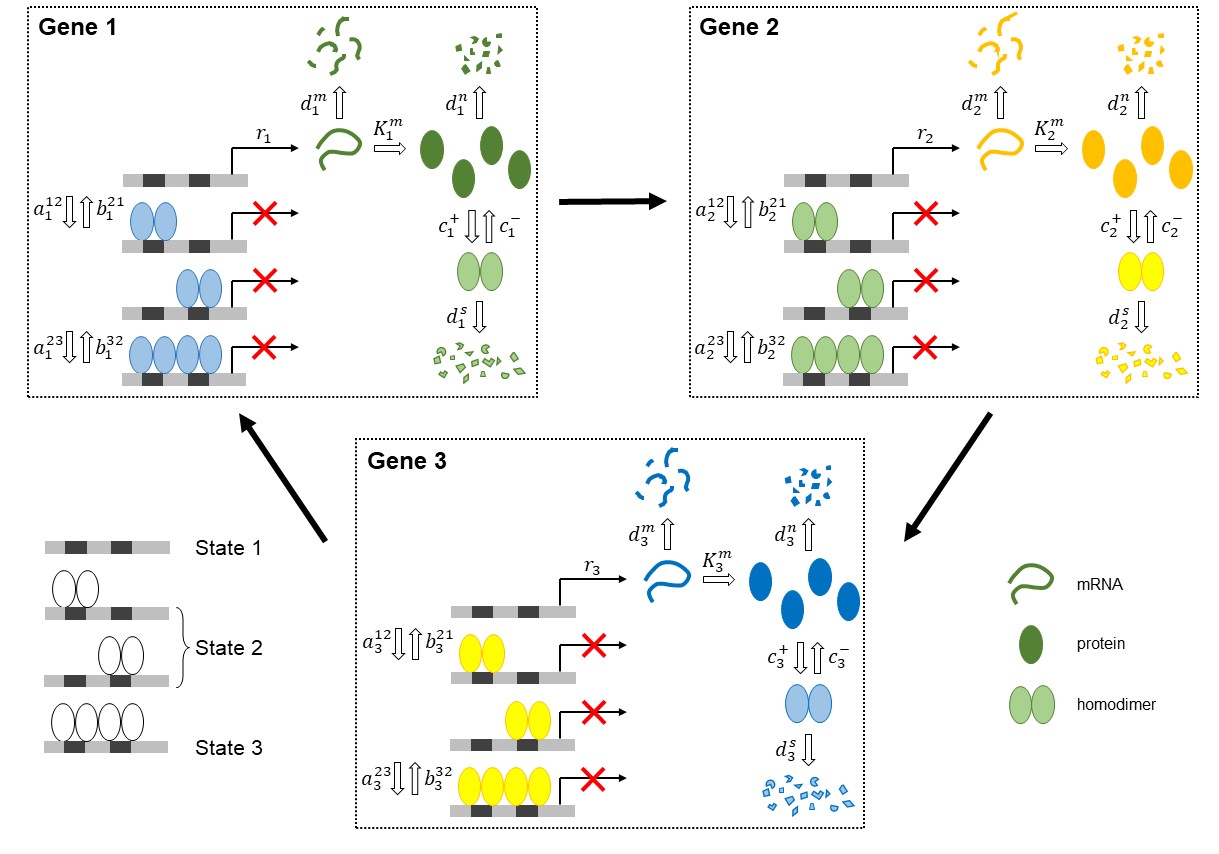}
\caption{A system of three interacting genes. The lower index of all parameters refers to the gene in question. 
The upper indices of $a_k^{ij}$ correspond to the transitions between promoter states (left to right); same notation applies to $b_k^{ij}$.
The parameters $r_k$ and $K_k^m$ give the transcription rate and translation rate, respectively, while in that same order,
$c_k^{+}$ and $c_k^{-}$ represent the rate of homodimer formation and dissociation.
Finally, $d_k^m$, $d_k^n$ and $d_k^s$ are the degradation rates of mRNA, protein and homodimer, respectively. 
}
\end{figure}

Let us partition the system as follows:
\begin{eqnarray}\label{reactionsex1}
&&\text{group $A$}\,\,\,\,\,\,\,\,\,\,\,\,\,\,\,\,\,\,\,\,\,\text{group $B$}\nonumber\\
&&1.\,\emptyset\longrightarrow m^k\,\,\,\,\,\,\,\,\,\,\,\,\,\,\,\,1.\,w^k_1+s^{k}\longrightarrow w^k_2\nonumber\\
&&2.\,m^k\longrightarrow \emptyset\,\,\,\,\,\,\,\,\,\,\,\,\,\,\,\,2.\,w^k_2+s^{k}\longrightarrow w^k_3\nonumber\\
&&\,\,\,\,\,\,\,\,\,\,\,\,\,\,\,\,\,\,\,\,\,\,\,\,\,\,\,\,\,\,\,\,\,\,\,\,\,\,\,\,\,\,\,\,\,3.\,w^k_2\longrightarrow w^k_1+s^{k}\nonumber\\
&&\,\,\,\,\,\,\,\,\,\,\,\,\,\,\,\,\,\,\,\,\,\,\,\,\,\,\,\,\,\,\,\,\,\,\,\,\,\,\,\,\,\,\,\,\,4.\,w^k_3\longrightarrow w^k_2+s^{k}\nonumber\\
&&\,\,\,\,\,\,\,\,\,\,\,\,\,\,\,\,\,\,\,\,\,\,\,\,\,\,\,\,\,\,\,\,\,\,\,\,\,\,\,\,\,\,\,\,\,5.\,m^k\longrightarrow m^k+n^k\nonumber\\
&&\,\,\,\,\,\,\,\,\,\,\,\,\,\,\,\,\,\,\,\,\,\,\,\,\,\,\,\,\,\,\,\,\,\,\,\,\,\,\,\,\,\,\,\,\,6.\,n^k\longrightarrow \emptyset\nonumber\\
&&\,\,\,\,\,\,\,\,\,\,\,\,\,\,\,\,\,\,\,\,\,\,\,\,\,\,\,\,\,\,\,\,\,\,\,\,\,\,\,\,\,\,\,\,\,7.\,2n^k\longrightarrow s^{k}\nonumber\\
&&\,\,\,\,\,\,\,\,\,\,\,\,\,\,\,\,\,\,\,\,\,\,\,\,\,\,\,\,\,\,\,\,\,\,\,\,\,\,\,\,\,\,\,\,\,8.\,s^{k}\longrightarrow 2n^k\nonumber\\
&&\,\,\,\,\,\,\,\,\,\,\,\,\,\,\,\,\,\,\,\,\,\,\,\,\,\,\,\,\,\,\,\,\,\,\,\,\,\,\,\,\,\,\,\,\,9.\,s^{k}\longrightarrow \emptyset,\nonumber\\
\end{eqnarray}
The propensities for $A$, $a_{\mu}(w^1,w^2,w^3)$, are given by
\begin{eqnarray}\label{PropensitiesEx1}
a_1&=&r_1\delta_{w^11}\nonumber\\
a_2&=&r_2\delta_{w^21}\nonumber\\
a_3&=&r_3\delta_{w^31}\nonumber\\
a_4&=&d^m_1m^1\nonumber\\
a_5&=&d^m_2m^2\nonumber\\
a_6&=&d^m_3m^3.
\end{eqnarray}
Before we deal with the modified CME (MCME) for $B$, let us first write down the CME:
\begin{eqnarray}\label{MEBa}
\frac{d{\bf P}}{dt}&=&{\bf W}({\bf s}){\bf P}+\sum_kK_km^k\left[{\bf P}(n^k-1)-{\bf P}\right]+d^n_k\left[(n^k+1){\bf P}(n^k+1)-n^k{\bf P}\right]\nonumber\\
&+&\sum_k\frac{c^{+}_{k}}{2}\left[(n^k+2)(n^k+1){\bf P}(n^k+2,s^{k}-1)-n^k(n^k+1){\bf P}\right]\nonumber\\
&+&\sum_kc^{-}_{k}\left[(s^{k}+1){\bf P}(n^k-2,s^{k}+1)-s^{k}{\bf P}\right]\nonumber\\
&+&\sum_kd^s_k\left[(s^k+1){\bf P}(s^k+1)-s^k{\bf P}\right],
\end{eqnarray}
where ${\bf P}$ is a vector whose dimension is equal to the number of all unique combinations of promoter states $(w^1,w^2,w^3)$ and
the elements of the matrix ${\bf W}({\bf s})$ give the propensities for the transitions between the combinations. 
We have employed a short hand notation in which ${\bf P}$ is short for ${\bf P}(n^1,n^2,n^3,s^1,s^2,s^3,t)$, 
${\bf P}(n^1+2,s^1-1)$ is short for ${\bf P}(n^1+2,n^2,n^3,s^1-1,s^2,s^3,t)$, etc. 

Note that reactions 1 to 4 in $B$ can be approximately decoupled from
reactions 5 to 9 on account of the fact that reactions 1 and 2 change the copy number of $s^{k}$ cyclically and at most by two: 
they can happen one after the other before the only available reaction
channel involving the promoter are reactions 3 and 4, which return one or both copies of $s^{k}$ to the system. 
Hence, if $s^{k}\gg2$ for all $k=1,2,3$, the CME for reactions 5 to 9 reads: 
\begin{eqnarray}\label{MEBns}
\frac{dP}{dt}&=&\sum_kK_km^k\left[P(n^k-1)-P\right]+d^n_k\left[(n^k+1)P(n^k+1)-n^kP\right]\nonumber\\
&+&\sum_k\frac{c^{+}_{k}}{2}\left[(n^k+2)(n^k+1)P(n^k+2,s^{k}-1)-n^k(n^k+1)P\right]\nonumber\\
&+&\sum_kc^{-}_{k}\left[(s^{k}+1)P(n^k-2,s^{k}+1)-s^{k}P\right]\nonumber\\
&+&\sum_kd^s_k\left[(s^k+1)P(s^k+1)-s^kP\right].
\end{eqnarray}
Although this equation cannot be solved exactly, it can be shown that the variances of $n^k$ and $s^{k}$ are very close to Poisson \cite{Pucci}. 
Thus, provided the averages $\langle n^k\rangle$ and $\langle s^k\rangle$ are much greater than 1, we can make
the approximation
\begin{equation}\label{aveapp}
\sum_{{\bf n},{\bf s}}f({\bf n},{\bf s})P({\bf n},{\bf s})=f(\langle{\bf n}\rangle,\langle{\bf s}\rangle),
\end{equation}
for any smooth function $f$. Thus, if we write the joint probability $P(l,{\bf n},{\bf s})$, where the index $l$
labels the elements of ${\bf P}$, as
\begin{equation}
P(l,{\bf n},{\bf s})=P(l|{\bf n},{\bf s})P({\bf n},{\bf s}),
\end{equation}
then, it follows from relation (\ref{aveapp}) that
\begin{eqnarray}
\sum_{{\bf n},{\bf s}}\sum_{q}W_{lq}({\bf s})P(q,{\bf n},{\bf s})&=&\sum_{{\bf n},{\bf s}}\left[\sum_{q}W_{lq}({\bf s})P(q|{\bf n},{\bf s})\right]P({\bf n},{\bf s})\nonumber\\
&=&\sum_{q}W_{lq}(\langle{\bf s}\rangle)P(q|\langle{\bf n}\rangle,\langle{\bf s}\rangle)=({\bf{\bar W}}{\bf P})_l,
\end{eqnarray}
where ${\bf \bar W}={\bf W}(\langle{\bf s}\rangle)$.
Hence, summing both sides of Eq. (\ref{MEBa}) over ${\bf n}$ and ${\bf s}$, we obtain
\begin{equation}\label{CMEprom}
\frac{d{\bf P}}{dt}={\bf{\bar W}}{\bf P}.
\end{equation}
The average $\langle s^{k}\rangle$ ($\langle n^{k}\rangle$) can be obtained by multiplying Eq. (\ref{MEBns}) by $s^{k}$ ($n^{k}$),
summing over ${\bf n}$ and ${\bf s}$, and remembering the relation (\ref{aveapp}):
\begin{eqnarray}\label{Averagesns}
\frac{d}{dt}\langle s^{k}\rangle&=&\frac{c^{+}_{k}}{2}\langle n^k\rangle(\langle n^k\rangle-1)-(c^{-}_{k}+d^s_k)\langle s^k\rangle\nonumber\\
\frac{d}{dt}\langle n^{k}\rangle&=&K_km^k+2c^{-}_{k}\langle s^k\rangle-c^{+}_{k}\langle n^k\rangle(\langle n^k\rangle-1)-d^n_k\langle n^k\rangle.
\end{eqnarray}
Since we are treating ${\bf n}$ and ${\bf s}$ as deterministic, the CME for the entire system $B$, is Eq. (\ref{CMEprom}).
For convenience, let us write it component by component:
\begin{equation}\label{CMEpromIndex}
\frac{d}{dt}P(w^1,w^2,w^3)=\sum_{w'^1,w'^2,w'^3}{\bar W}(w^1,w^2,w^3;w'^1,w'^2,w'^3)P(w'^1,w'^2,w'^3).
\end{equation}
Since we know that a promoter state of one gene does not influence a promoter state of another gene, $P(w^1,w^2,w^3)$ must be a product,
$P_1(w^1)P_2(w^2)P_3(w^3)$, such that $P_k(w^k)$ satisfies
\begin{equation}\label{CMEprom_k}
\frac{d{\bf P}_k}{dt}={\bf M}_k{\bf P}_k,
\end{equation}
where
\begin{eqnarray}\label{M}
{\bf M}_k=
\left[\begin{array}{cccccccccc}
-2a_k\langle s^{k}\rangle & b_k & 0 \\
2a_k\langle s^{k}\rangle & -a_k\langle s^{k}\rangle-b_k & 2b_k \\
0 & a_k\langle s^{k}\rangle  & -2b_k \nonumber
\end{array}\right].
\end{eqnarray}
Here, $a_k=a_k^{12}=a_k^{23}$ and $b_k=b_k^{21}=b_k^{32}$.
Taking a derivative of $P_1(w^1)P_2(w^2)P_3(w^3)$, we obtain
\begin{eqnarray}
\frac{d}{dt}P(w^1,w^2,w^3)&=&
\sum_{w'^1}M_1(w^1,w'^1)P(w'^1,w^2,w^3)+\sum_{w'^2}M_2(w^2,w'^2)P(w^1,w'^2,w^3)\nonumber\\
&+&\sum_{w'^3}M_3(w^3,w'^3)P(w^1,w^2,w'^3).
\end{eqnarray}
Following Eq. (\ref{ModME}), the MCME acquires the form
\begin{eqnarray}
\frac{d}{dt}Q(w^1,w^2,w^3)&=&
\sum_{w'^1}M_1(w^1,w'^1)Q(w'^1,w^2,w^3)+\sum_{w'^2}M_2(w^2,w'^2)Q(w^1,w'^2,w^3)\nonumber\\
&+&\sum_{w'^3}M_3(w^3,w'^3)Q(w^1,w^2,w'^3)\nonumber\\
&-&\sum_i(r_i\delta_{w^i,1}+d^m_im^i)Q(w^1,w^2,w^3).
\end{eqnarray}
We can simplify this equation by defining a new function ${\tilde Q}(w^1,w^2,w^3)$, such that
\begin{equation}
Q(w^1,w^2,w^3)=\text{exp}\left[-\sum_id^m_im^it\right]{\tilde Q}(w^1,w^2,w^3).
\end{equation}
Hence,
\begin{eqnarray}\label{MCMEQbar}
\frac{d}{dt}{\tilde Q}(w^1,w^2,w^3)&=&
\sum_{w'^1}M_1(w^1,w'^1){\tilde Q}(w'^1,w^2,w^3)+\sum_{w'^2}M_2(w^2,w'^2){\tilde Q}(w^1,w'^2,w^3)\nonumber\\
&+&\sum_{w'^3}M_3(w^3,w'^3){\tilde Q}(w^1,w^2,w'^3)\nonumber\\
&-&\sum_ir_i\delta_{w^i,1}{\tilde Q}(w^1,w^2,w^3).
\end{eqnarray}
Since the initial conditions demand that $Q(w^1,w^2,w^3,t=0)={\tilde Q}(w^1,w^2,w^3,t=0)=P_{in,1}(w^1,0)P_{in,2}(w^2,0)P_{in,3}(w^3,0)$,
${\tilde Q}(w^1,w^2,w^3)$ can also be written as a product ${\tilde Q}_1(w^1){\tilde Q}_2(w^2){\tilde Q}_3(w^3)$, which, when inserted
into Eq. (\ref{MCMEQbar}) yields
\begin{eqnarray}\label{QQQ}
&&{\tilde Q}_2(w^2){\tilde Q}_3(w^3)\left[\frac{d}{dt}{\tilde Q}_1(w^1)-\sum_{w'^1}M_1^{w^1,w'^1}{\tilde Q}_1(w'^1)+r_1\delta_{w^1,1}{\tilde Q}_1(w^1)\right]+\nonumber\\
&&{\tilde Q}_1(w^1){\tilde Q}_3(w^3)\left[\frac{d}{dt}{\tilde Q}_2(w^2)-\sum_{w'^2}M_2^{w^2,w'^2}{\tilde Q}_1(w'^2)+r_2\delta_{w^2,1}{\tilde Q}_2(w^2)\right]+\nonumber\\
&&{\tilde Q}_1(w^1){\tilde Q}_2(w^2)\left[\frac{d}{dt}{\tilde Q}_3(w^3)-\sum_{w'^3}M_3^{w^3,w'^3}{\tilde Q}_3(w'^3)+r_3\delta_{w^3,1}{\tilde Q}_3(w^3)\right]=0.
\end{eqnarray}
Thus, our $Q$-distribution is given by
\begin{equation}\label{Qdistpath}
Q(w^1,w^2,w^3,t)=\text{exp}\left[-\sum_id^m_im^it\right]{\tilde Q}_1(w^1,t){\tilde Q}_2(w^2,t){\tilde Q}_3(w^3,t),
\end{equation}
with ${\tilde Q}_k(w^k)$ satisfying
\begin{equation}\label{QdistEx1}
\frac{d}{dt}{\bf {\tilde Q}}_k=[{\bf M}_k-{\bf R}_k]{\bf {\tilde Q}}_k,
\end{equation}
where 
\begin{eqnarray}\label{R}
{\bf R}_k=
\left[\begin{array}{cccccccccc}
r_k & 0 & 0 \\
0 & 0 & 0 \\
0 & 0 & 0 \nonumber
\end{array}\right].
\end{eqnarray}

We are now ready to compute $Q(t)$, $Q(\nu|t)$ and $Q({\bf z}_t|,\mu,\tau)$ (step 3 of the HSSA).
The first one is simply
\begin{equation}\label{qQtau}
Q(t)=\text{exp}\left[-\sum_id^m_im^it\right]{\tilde Q}_1(t){\tilde Q}_2(t){\tilde Q}_3(t),
\end{equation}
where
\begin{equation}
{\tilde Q}_k(t)=\sum_{w^k}{\tilde Q}_k(w^k,t);
\end{equation}
the second one reads
\begin{equation}\label{qQmutau}
Q(\mu|\tau)=\sum_{w^1w^2w^3}{\tilde Q}_1(w^1,\tau){\tilde Q}_2(w^2,\tau){\tilde Q}_3(w^3,\tau)a_{\mu}(w^1,w^2,w^3)/{\cal Q}(\tau),
\end{equation}
where
\begin{equation}
{\cal Q}(\tau)=\sum_{\mu}\sum_{w^1w^2w^3}{\tilde Q}_2(w^2,\tau){\tilde Q}_3(w^3,\tau)a_{\mu}(w^1,w^2,w^3);
\end{equation}
while the third one is given by
\begin{equation}\label{qQzmutau}
Q(w^1,w^2,w^3|\mu,\tau)={\tilde Q}_2(w^2,\tau){\tilde Q}_3(w^3,\tau)a_{\mu}(w^1,w^2,w^3)/
{\cal Q}(\tau).
\end{equation}
Inserting the propensities listed in Eq. (\ref{PropensitiesEx1}) into (\ref{qQmutau}), we obtain
\begin{eqnarray}\label{choosemu}
Q(1|\tau)&=&\frac{r_1{\tilde Q}_1(1,\tau){\tilde Q}_2(\tau){\tilde Q}_3(\tau)}{{\cal Q}(\tau)}\nonumber\\
Q(2|\tau)&=&\frac{r_2{\tilde Q}_1(\tau){\tilde Q}_2(1,\tau){\tilde Q}_3(\tau)}{{\cal Q}(\tau)}\nonumber\\
Q(3|\tau)&=&\frac{r_3{\tilde Q}_1(\tau){\tilde Q}_2(\tau){\tilde Q}_3(1,\tau)}{{\cal Q}(\tau)}\nonumber\\
Q(4|\tau)&=&\frac{d^m_1m^1{\tilde Q}_1(\tau){\tilde Q}_2(\tau){\tilde Q}_3(\tau)}{{\cal Q}(\tau)}\nonumber\\
Q(5|\tau)&=&\frac{d^m_2m^2{\tilde Q}_1(\tau){\tilde Q}_2(\tau){\tilde Q}_3(\tau)}{{\cal Q}(\tau)}\nonumber\\
Q(6|\tau)&=&\frac{d^m_3m^3{\tilde Q}_1(\tau){\tilde Q}_2(\tau){\tilde Q}_3(\tau)}{{\cal Q}(\tau)}\nonumber\\
\end{eqnarray}
Similarly, if we insert the explicit reaction propensities into (\ref{qQzmutau}), we obtain the updates for the initial probability distributions,
$P_{\text{in},k}(w^k,0)=Q_k(w^k,\tau)$:
\begin{eqnarray}\label{updateP}
&&\mu=1:\,\,\,\,\,\,Q(w^1|1,\tau)=\delta_{w^11},\,\,\,\,\,\,Q(w^2|1,\tau)=\frac{{\tilde Q}_2(w^2,\tau)}{{\tilde Q}_2(\tau)},\,\,\,\,\,\,Q(w^3|1,\tau)=\frac{{\tilde Q}_3(w^3,\tau)}{{\tilde Q}_3(\tau)},\nonumber\\
&&\mu=2:\,\,\,\,\,\,Q(w^2|2,\tau)=\delta_{w^21},\,\,\,\,\,\,Q(w^1|2,\tau)=\frac{{\tilde Q}_1(w^1,\tau)}{{\tilde Q}_2(\tau)},\,\,\,\,\,\,Q(w^3|2,\tau)=\frac{{\tilde Q}_3(w^3,\tau)}{{\tilde Q}_3(\tau)},\nonumber\\
&&\mu=3:\,\,\,\,\,\,Q(w^3|3,\tau)=\delta_{w^31},\,\,\,\,\,\,Q(w^1|3,\tau)=\frac{{\tilde Q}_1(w^1,\tau)}{{\tilde Q}_2(\tau)},\,\,\,\,\,\,Q(w^2|3,\tau)=\frac{{\tilde Q}_2(w^2,\tau)}{{\tilde Q}_2(\tau)},\nonumber\\
&&\mu=4,5,6:\nonumber\\
&&\,\,\,\,\,\,Q(w^1|\mu,\tau)=\frac{{\tilde Q}_1(w^1,\tau)}{{\tilde Q}_1(\tau)},\,\,\,\,\,\,Q(w^2|\mu,\tau)=\frac{{\tilde Q}_2(w^2,\tau)}{{\tilde Q}_2(\tau)},\,\,\,\,\,\,Q(w^3|\mu,\tau)=\frac{{\tilde Q}_3(w^3,\tau)}{{\tilde Q}_3(\tau)}
\end{eqnarray}
Since none of the reactions in $A$ change either of the copy numbers ${\bf n}$ or ${\bf s}$, the shifting operator ${\hat S}_\mu$ is just unity.

In principle, we could now run the HSSA as prescribed in the previous section. However, in order to make it as efficient as possible, we must find a way
to deal with Eqs. (\ref{Averagesns}) and (\ref{QdistEx1}). Although in general Eqs. (\ref{Averagesns}) cannot be solved analytically, we can exploit the fact that $\tau$ will always be small compared to
some characteristic time during which $\langle n^k\rangle$ and $\langle s^k\rangle$ will change significantly. For such $\tau$ we can write $\langle n^k\rangle$ and $\langle s^k\rangle$  
as polynomials on $\tau$:
\begin{equation}\label{polynom}
\langle n^k\rangle=\sum_{q=0}u^k_{q}\tau^q\,\,\,\,\,\,\,\langle s^k\rangle=\sum_{q=0}v^k_{q}\tau^q.
\end{equation}
Plugging these into Eqs. (\ref{Averagesns}), we obtain algebraic equations for $u^k_{q}$ and $v^k_{q}$, which, up to order $\tau^2$, read 
\begin{eqnarray}
v^k_{1}&=&\frac{c^{+}_{k}}{2}u^k_{0}(u^k_{0}-1)-(c^{-}_{k}+d^s_k)v^k_{0}\nonumber\\
u^k_{1}&=&K_km^k+2c^{-}_{k}v^k_{0}-c^{+}_{k}u^k_{0}(u^k_{0}-1)-d^n_ku^k_{0}\nonumber\\
2v^k_{2}&=&\frac{c^{+}_{k}}{2}(2u^k_{0}u^k_{1}-u^k_{1})-(c^{-}_{k}+d^s_k)v^k_{1}\nonumber\\
2u^k_{2}&=&2c^{-}_{k}v^k_{1}-c^{+}_{k}(2u^k_{0}u^k_{1}-u^k_{1})-d^n_ku^k_{1}.
\end{eqnarray}
The smallness of $\tau$ can also be useful for simplifying Eq. (\ref{QdistEx1}). If we write the matrix ${\bf M}_k$ as
${\bf M}_k={\bf M}^{(0)}_k+{\boldsymbol\Delta}_k(t)$, where
\begin{eqnarray}
{\bf M}^{(0)}_k=
\left[\begin{array}{cccccccccc}
-2a_ku^k_0 & b_k & 0 \\
2a_ku^k_0 & -a_ku^k_0-b_k & 2b_k \\
0 & a_ku^k_0 & -2b_k \nonumber
\end{array}\right],
\end{eqnarray}
and
\begin{eqnarray}
{\boldsymbol\Delta}_k(t)=\sum_{q=1}a_ku^k_qt^q
\left[\begin{array}{cccccccccc}
-2 & 0 & 0 \\
2 & -1 & 0 \\
0 & 1 & 0 \nonumber
\end{array}\right],
\end{eqnarray}
Eq. (\ref{QdistEx1}) becomes
\begin{equation}\label{QdistEx1simple}
\frac{d}{dt}{\bf {\tilde Q}}_k=[({\bf M}^{(0)}_k-{\bf R}_k)+{\boldsymbol\Delta}_k(t)]{\bf {\tilde Q}}_k.
\end{equation}
Since ${\boldsymbol\Delta}_k(t)$ is a correction to the matrix ${\bf M}^{(0)}_k-{\bf R}_k$, we can
write ${\bf {\tilde Q}}_k={\bf {\tilde Q}}_k^{(0)}+{\bf {\tilde Q}}_k^{(1)}+...$, where the superscript
indicates the order on ${\boldsymbol\Delta}_k(t)$, i. e. ${\bf {\tilde Q}}_k^{(n)}\sim{\cal O}({\boldsymbol\Delta}_k(t)^n)$.
Plugging ${\bf {\tilde Q}}_k$ thus expanded into Eq. (\ref{QdistEx1simple}), and collecting terms of the same order
on ${\boldsymbol\Delta}_k(t)$, we obtain this series of equations:
\begin{eqnarray}
\frac{d}{dt}{\bf {\tilde Q}}_k^{(0)}&=&({\bf M}^{(0)}_k-{\bf R}_k){\bf {\tilde Q}}_k^{(0)}\nonumber\\
\frac{d}{dt}{\bf {\tilde Q}}_k^{(n)}&=&({\bf M}^{(0)}_k-{\bf R}_k){\bf {\tilde Q}}_k^{(n)}+{\boldsymbol\Delta}_k(t){\bf {\tilde Q}}_k^{(n-1)}\,\,\,\,\,\forall n>0.
\end{eqnarray}
These equations can be solved analytically for an arbitrary $n$ in terms of the solution to ${\bf {\tilde Q}}_k^{(0)}$, which is
\begin{equation}\label{Qsol}
{\bf {\tilde Q}}_k^{(0)}(t)=\sum_pe^{E^k_pt}\left({\bf U}_k{\bf B}_p{\bf U}_k^{-1}\right){\bf P}_{\text{in},k},
\end{equation}
where $E^k_p$ and ${\bf U}_k$ are the $p^{\text{th}}$ eigenvalue and the column eigenvectors of ${\bf M}^{(0)}_k-{\bf R}_k$, respectivelly, and
\begin{eqnarray}
{\bf B}_p=
\left[\begin{array}{cccccccccc}
\delta_{p1} & 0 & 0 \\
0 & \delta_{p2} & 0 \\
0 & 0 & \delta_{p3} \nonumber
\end{array}\right].
\end{eqnarray}
For the purpose of testing, we will consider Eq. (\ref{Qsol}) to be our $Q$-distribution. Plugging Eq. (\ref{Qsol}) into
Eq. (\ref{qQtau}), we obtain
\begin{equation}\label{qQtauFin}
Q(t)=\text{exp}\left[-\sum_id^m_it\right]\left(\text{tr}{\bf {\tilde Q}}_1^{(0)}(t)\right)
\left(\text{tr}{\bf {\tilde Q}}_2^{(0)}(t)\right)\left(\text{tr}{\bf {\tilde Q}}_3^{(0)}(t)\right).
\end{equation}
$\tau$ can now be computed by choosing a random real number $\xi=(0,1]$ and solving $Q(\tau)=\xi$ for $\tau$.
Once we have a value of $\tau$, expressions in (\ref{choosemu}) and (\ref{updateP}) can be computed from this relation:
\begin{equation}
{\tilde Q}_k(w^k,\tau)=({\bf e}_{w^k})^T{\bf {\tilde Q}}_k^{(0)}(\tau),
\end{equation}
where
\begin{eqnarray}
{\bf e}_i=
\left[\begin{array}{ccc}
\delta_{i1}\\
\delta_{i2}\\
\delta_{i3} \nonumber
\end{array}\right].
\end{eqnarray}

In order to speed up the numerical search for $\tau$
we first computed the average $\tau$,
\begin{equation}
\tau_{Av}=\int_0^\infty dtQ(t),
\end{equation}
and then evaluated $Q(\tau)-\xi$ for $\tau=\tau_{Av}n/10$ for $n=0,1,2,...$ and stopped when $Q(\tau_{Av}n/10)-\xi$ became negative.
Finally, we passed a straight line $y=mt+b$ through the points $Q(\tau_{Av}(n-1)/10)$ and $Q(\tau_{Av}n/10)$, and set $\tau=(\xi-b)/m$.
The accuracy and speed gain relative to the GA are discussed in the results section.

\subsection{Example 2}

In this example we choose a less ambitious partitioning of the system:
\begin{eqnarray}\label{reactionsEx2}
&&\text{group $A$}\,\,\,\,\,\,\,\,\,\,\,\,\,\,\,\,\,\,\,\,\,\,\,\,\,\,\,\,\,\,\,\,\,\,\,\,\,\,\,\,\,\,\text{group $B$}\nonumber\\
&&\emptyset\longrightarrow m^k\,\,\,\,\,\,\,\,\,\,\,\,\,\,\,\,\,\,\,\,\,\,\,\,\,\,\,\,\,\,\,\,\,\,\,\,\,\,\,m^k\longrightarrow m^k+n^k\nonumber\\
&&m^k\longrightarrow \emptyset\,\,\,\,\,\,\,\,\,\,\,\,\,\,\,\,\,\,\,\,\,\,\,\,\,\,\,\,\,\,\,\,\,\,\,\,\,\,\,n^k\longrightarrow \emptyset\nonumber\\
&&w^k_1+s^{k}\longrightarrow w^k_2\,\,\,\,\,\,\,\,\,\,\,\,\,\,\,\,\,\,\,\,\,\,\,2n^k\longrightarrow s^{k}\nonumber\\
&&w^k_2+s^{k}\longrightarrow w^k_3\,\,\,\,\,\,\,\,\,\,\,\,\,\,\,\,\,\,\,\,\,\,\,s^{k}\longrightarrow 2n^k\nonumber\\
&&w^k_2\longrightarrow w^k_1+s^{k}\,\,\,\,\,\,\,\,\,\,\,\,\,\,\,\,\,\,\,\,\,\,\,s^{k}\longrightarrow \emptyset\nonumber\\
&&w^k_3\longrightarrow w^k_2+s^{k}.\nonumber\\
\end{eqnarray}
The propensities for $A$ are the same as in the previous example, but in addition we have
\begin{eqnarray}\label{PropensitiesEx2}
a_7&=&a_1^{12}w_1^1s^3\nonumber\\
a_8&=&a_1^{23}w_2^1s^3\nonumber\\
a_9&=&b_1^{21}w_2^1\nonumber\\
a_{10}&=&b_1^{32}w_3^1\nonumber\\
a_{11}&=&a_2^{12}w_1^2s^1\nonumber\\
a_{12}&=&a_2^{23}w_2^2s^1\nonumber\\
a_{13}&=&b_2^{21}w_2^2\nonumber\\
a_{14}&=&b_2^{32}w_3^2\nonumber\\
a_{15}&=&a_3^{12}w_1^3s^2\nonumber\\
a_{16}&=&a_3^{23}w_2^3s^2\nonumber\\
a_{17}&=&b_3^{21}w_2^3\nonumber\\
a_{18}&=&b_3^{32}w_3^3.
\end{eqnarray}
The CME for $B$ is the same as Eq. (\ref{MEBns}),
while the MCME reads
\begin{equation}\label{MEBnsEx2}
\frac{dQ}{dt}={\cal H}(Q)
-\left[\sum_{i=1}^3(r_i\delta_{w^i,1}+d^m_im^i)+\sum_{i=1}^3\sum_{j=1}^3(a_i^js^{i-1}+b_i^j)w_i^j\right]Q,
\end{equation}
where, for simplicity of notation, we made the definition
\begin{eqnarray}\label{MEBnsEx2}
{\cal H}(Q)&=&\sum_kK_km^k\left[Q(n^k-1)-Q\right]+d^n_k\left[(n^k+1)Q(n^k+1)-n^kQ\right]\nonumber\\
&+&\sum_k\frac{c^{+}_{k}}{2}\left[(n^k+2)(n^k+1)Q(n^k+2,s^{k}-1)-n^k(n^k+1)Q\right]\nonumber\\
&+&\sum_kc^{-}_{k}\left[(s^{k}+1)Q(n^k-2,s^{k}+1)-s^{k}Q\right]\nonumber\\
&+&\sum_kd^s_k\left[(n^k+1)Q(n^k+1)-n^kQ\right]
\end{eqnarray}
Defining a new variable ${\tilde Q}$, such that
\begin{eqnarray}\label{BbarEx2}
Q({\bf n},{\bf s},t)&=&{\tilde Q}({\bf n},{\bf s},t)\nonumber\\
&\times&\text{exp}\left\{-\left[\sum_{i=1}^3(r_i\delta_{w^i,1}+d^m_im^i)+\sum_{j=1}^3\sum_{j=1}^3b_i^jw_i^j)\right]t\right\}\nonumber\\
&\times&\text{exp}\left\{-\sum_{j=1}^3\sum_{j=1}^3\int_0^tdt'a_i^j\langle s^{i-1}\rangle(t')\right\},
\end{eqnarray}
where 
\begin{equation}
\langle s^{i}\rangle(t)=\sum_{{\bf n},{\bf s}}s^{i}P({\bf n},{\bf s},t),
\end{equation}
and inserting (\ref{BbarEx2}) into Eq. (\ref{MEBnsEx2}), we obtain
\begin{eqnarray}\label{QbarEx2}
\frac{d{\tilde Q}}{dt}={\cal H}({\tilde Q})
-\Delta(t){\tilde Q},
\end{eqnarray}
where
\begin{equation}
\Delta(t)=\sum_{i=1}^3\sum_{j=1}^3a_i^j(s^{i-1}-\langle s^{i-1}\rangle).
\end{equation}
As before, we will treat $\Delta(t)$ as small and write ${\tilde Q}={\tilde Q}^{(0)}+{\tilde Q}^{(1)}+...$
where ${\tilde Q}^{(n)}\sim{\cal O}(\Delta(t)^n)$, and insert it into Eq. (\ref{QbarEx2}) to obtain
\begin{eqnarray}\label{QbarDelta}
\frac{d{\tilde Q}^{(0)}}{dt}&=&{\cal H}({\tilde Q}^{(0)})\nonumber\\
\frac{d{\tilde Q}^{(n)}}{dt}&=&{\cal H}({\tilde Q}^{(n)})-\Delta(t){\tilde Q}^{(n-1)}\,\,\,\,\,\forall n>0.
\end{eqnarray}
The first equation is identical to the CME in (\ref{MEBns}), and since the initial conditions must satisfy the relation
${\tilde Q}^{(0)}({\bf n},{\bf s},0)=P_{\text{in}}({\bf n},{\bf s},0)$, ${\tilde Q}^{(0)}$ is the probability
$P({\bf n},{\bf s},t)$. If we sum the second equation for $n=1$ over ${\bf n}$ and ${\bf s}$, we obtain
\begin{eqnarray}
\frac{d{\tilde Q}^{(1)}}{dt}&=&0,
\end{eqnarray}
which implies that ${\tilde Q}^{(1)}({\bf n},{\bf s},t)=0$. Hence, we can write
\begin{eqnarray}\label{BbarEx2Sum}
Q(t)&=&\sum_{{\bf n},{\bf s}}Q({\bf n},{\bf s},t)\nonumber\\
&=&\left[1+{\tilde Q}^{(2)}(t)+...\right]\nonumber\\
&\times&\text{exp}\left\{-\left[\sum_{i=1}^3(r_i\delta_{w^i,1}+d^m_im^i)+\sum_{j=1}^3\sum_{j=1}^3b_i^jw_i^j)\right]t\right\}\nonumber\\
&\times&\text{exp}\left\{-\sum_{j=1}^3\sum_{j=1}^3\int_0^tdt'a_i^j\langle s^{i-1}\rangle(t')\right\}.
\end{eqnarray}
For the purpose of testing, we set $Q^{(n>1)}(t)=0$.
As in the previous example, we expand the averages $\langle n^k\rangle$ and $\langle s^k\rangle$
in powers of $\tau$ and keep only the lowest terms, which yields 
\begin{equation}
Q(\tau)=e^{-R\tau},
\end{equation}
where
\begin{eqnarray}\label{BbarEx2Sum}
R=\sum_{i=1}^3(r_i\delta_{w^i,1}+d^m_im^i)+\sum_{j=1}^3\sum_{j=1}^3(a_i^jv_k^{(0)}+b_i^jw_i^j).
\end{eqnarray}
In this case, computing $\tau$ from $Q(\tau)-\xi_1=0$ is particularly simple:
\begin{equation}
\tau=-\frac{1}{R}\ln\xi_1.
\end{equation}
To compute the probability for reaction $\mu$ to occur once we sample $\tau$, we simply follow the
prescription in Eq. (\ref{Qmut}):
\begin{equation}\label{QmutEx2}
Q(\mu|\tau)=\frac{\sum_{{\bf n},{\bf s}}{\tilde Q}^{(0)}({\bf n},{\bf s},\tau)a_{\mu}({\bf n},{\bf s})}
{\sum_{\mu}\sum_{{\bf n},{\bf s}}{\tilde Q}^{(0)}({\bf n},{\bf s},\tau)a_{\mu}({\bf n},{\bf s})}
=\frac{a_{\mu}({\bf \bar n}(\tau),{\bf \bar s}(\tau))}{\sum_{\mu}a_{\mu}({\bf \bar n}(\tau),{\bf \bar s}(\tau))},
\end{equation}
where ${\bf \bar n}(\tau)=\langle{\bf n}\rangle=\sum_q{\bf u}^{(q)}\tau^q$ and ${\bf \bar s}(\tau)=\langle{\bf s}\rangle=\sum_q{\bf v}^{(q)}\tau^q$.
To update the probabilities, we follow Eq. (\ref{Qzmut}):
\begin{equation}
Q({\bf n},{\bf s}|\mu,\tau)=\frac{{\tilde Q}^{(0)}({\bf n},{\bf s},\tau)a_{\mu}({\bf n},{\bf s})}
{\sum_{{\bf n},{\bf s}}{\tilde Q}^{(0)}({\bf n},{\bf s},\tau)a_{\mu}({\bf n},{\bf s})}=\frac{{\tilde Q}^{(0)}({\bf n},{\bf s},\tau)a_{\mu}({\bf n},{\bf s})}
{a_{\mu}(\langle{\bf n}\rangle,\langle{\bf s}\rangle)}.
\end{equation}
By virtue of the relation (\ref{aveapp}), the updated averages $\langle{\bf n}\rangle$ and $\langle{\bf s}\rangle$
become
\begin{eqnarray}
{\bf n}&=&{\hat S}_\mu\frac{\sum_{{\bf n},{\bf s}}{\tilde Q}^{(0)}({\bf n},{\bf s},\tau){\bf n}a_{\mu}({\bf n},{\bf s})}
{a_{\mu}(\langle{\bf n}\rangle,\langle{\bf s}\rangle)}={\hat S}_\mu{\bf n}\nonumber\\
{\bf s}&=&{\hat S}_\mu\frac{\sum_{{\bf n},{\bf s}}{\tilde Q}^{(0)}({\bf n},{\bf s},\tau){\bf s}a_{\mu}({\bf n},{\bf s})}
{a_{\mu}(\langle{\bf n}\rangle,\langle{\bf s}\rangle)}={\hat S}_\mu{\bf s}.
\end{eqnarray}
We can approximate the effect of ${\hat S}_\mu$ by letting
\begin{eqnarray}
\left[\begin{array}{cc}
{\bf \bar n}(0)\\
{\bf \bar s}(0) \nonumber
\end{array}\right]
\,\,\,\rightarrow\,\,
\left[\begin{array}{cc}
{\bf \bar n}(\tau)\\
{\bf \bar s}(\tau) \nonumber
\end{array}\right]+{\bf h}_{\mu},
\end{eqnarray}
for those ${\bar n}^k$ and ${\bar s}^k$ that update to a value greater than or equal to zero;
those that update to a negative value are left unchanged.

\section{Results}

\begin{figure}[H]
\centering
\includegraphics[trim=0 0 0 1.0cm, height=0.4\textheight]{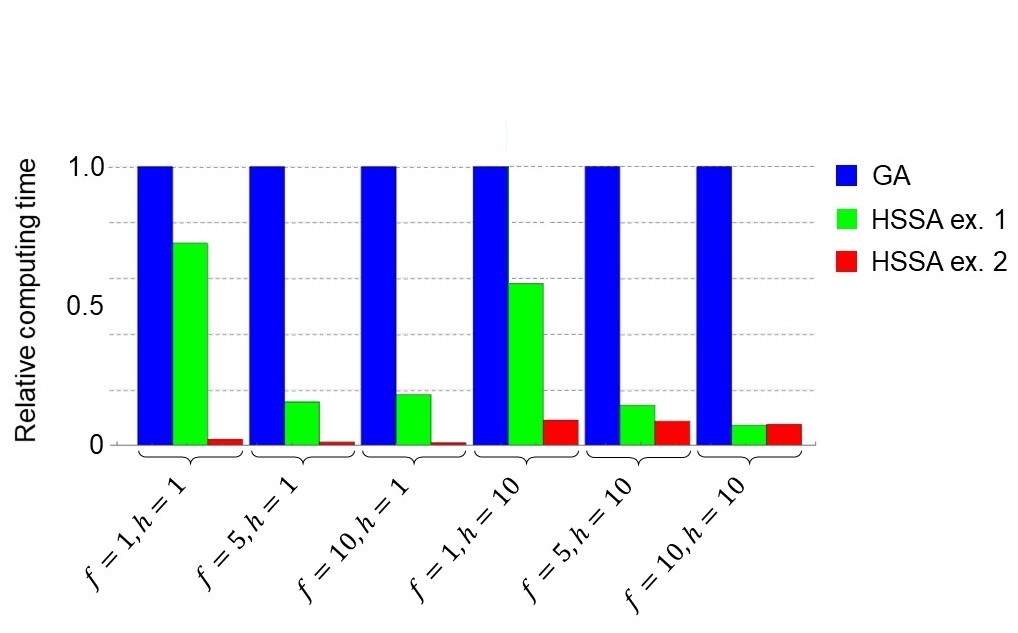}
\caption{Computing times for HSSAex1 and HSSAex2 relative to the computing time of GA (normalized to one), for
different control parameters $f$ and $h$. The computing times were averaged over 100 realizations running from 0 to 33 hours.}
\end{figure}

\begin{figure}[H]
\centering
\includegraphics[trim=0 0 0 1.0cm, height=0.8\textheight]{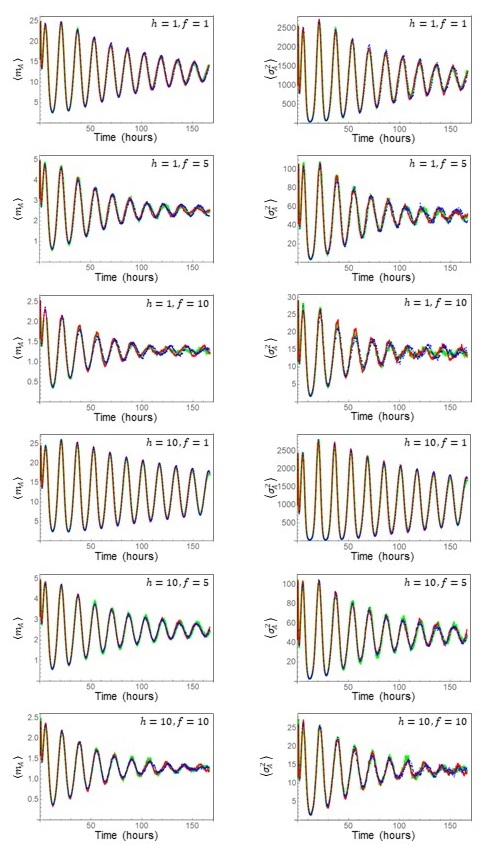}
\caption{Average and variance of $m^1$ as a function of time and different control parameters $f$ and $h$
for the GA (blue), HSSAex1 (green) and HSSAex2 (red). The ensemble size
was 500.}
\end{figure}

\begin{figure}[H]
\centering
\includegraphics[trim=0 0 0 1.0cm, height=0.6\textheight]{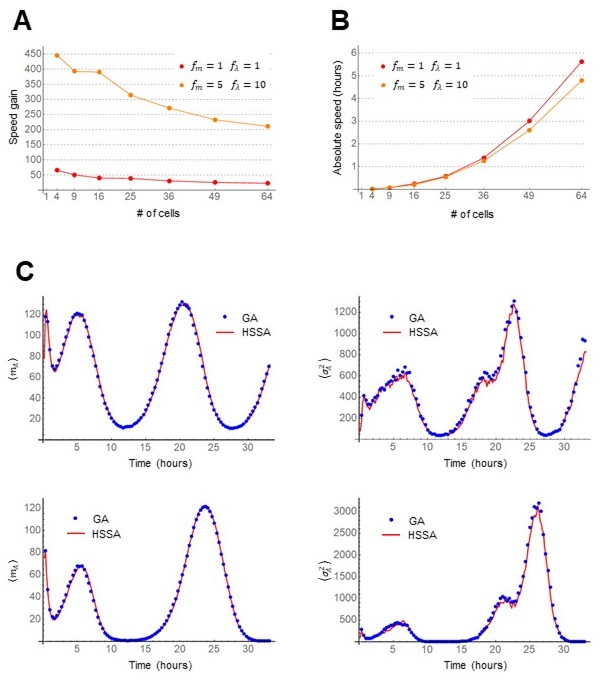}
\caption{A) Speed gain relative to the GA for HSSAex2 as a function of the control parameters $f_m$ and
$f_\lambda$ and number of cells. B) Absolute computing time as a function of the control parameters $f_m$ and
$f_\lambda$ and number of cells. C) Average and variance of $m^1$ belonging to the first cell as a function of time $f_m=f_\lambda=1$. The ensemble size
was 500. For A) and B), the simulations were performed up to 166 hours.}
\end{figure}

We have simulated the system described in Fig. 1 with the GA, and the two variations of the HSSA detailed in examples 1 and 2, which we label as
HSSAex1 and HSSAex2. The parameters
were chosen so as to make the system oscillatory:

\begin{eqnarray}
&&a_1^{1,2}=a_1^{2,3}=a_2^{1,2}=a_2^{2,3}=h\times10^{-3}\text{min}^{-1}\,\,\,\,\,a_3^{1,2}=a_3^{2,3}=h\times4\times10^{-3}\text{min}^{-1}\nonumber\\
&&b_1^{2,1}=b_1^{3,2}=b_2^{2,1}=b_2^{3,2}=b_3^{2,1}=b_3^{3,2}=h^{-1}\text{min}^{-1}\nonumber\\
&&r_1=r_2=r_3=10\times f^{-1}\text{min}^{-1}\nonumber\\
&&K^m_1=K^m_2=K^m_3=f\,\,\text{min}^{-1}\nonumber\\
&&d^m_1=d^m_2=d^m_3=5\times 10^{-2}\text{min}^{-1}\nonumber\\
&&d^n_1=d^n_2=d^n_3=1\times 10^{-2}\text{min}^{-1}\nonumber\\
&&d^s_1=d^s_2=d^s_3=5\times 10^{-3}\text{min}^{-1}\nonumber\\
&&c^+_1=c^+_2=c^+_3=5\times 10^{-5}\text{min}^{-1}\nonumber\\
&&c^-_1=c^-_2=c^-_3=5\times 10^{-2}\text{min}^{-1}.
\end{eqnarray}
The parameters $h$ and $f$ allow us to change the system without loosing oscillations.
Fig. 2 shows the efficiency of HSSAex1 and HSSAex2 relative to the GA, which has been normalized to one for convenience.
The green column represents the computing time of HSSAex1 relative to the computing time of the GA; the red column gives the
relative computing time of HSSAex2. The HSSAex2 is clearly more efficient compared
to HSSAex2 in all but the last case ($f=h=10$), but even in this case they are very close.

Regarding accuracy, Fig. 3 shows the average and variance of the mRNA copy number of the first gene for the GA (blue),
HSSAex1 (green) and HSSAex2 (red) for the 6 cases shown in Fig. 2. Although the efficiency between the HSSAex1 and HSSAex2 may differ greatly,
their accuracy is equally good.

Given the impressive speed gain of HSSAex2, we decided to test it on a 2-dimensional array of coupled identical cells. The
coupling was introduced via a simple diffusion process of the homodimer of gene 3: $s^3_j\rightarrow  s^3_{j'}$, where
$j$ and $j'$ label two adjacent cells. The diffusion coefficient was chosen to be $\lambda\gamma_{\lambda}$, where $\lambda=10^{-2}$ min$^{-1}$
and $\gamma_{\lambda}$ is a dimensionless control parameter. Another control parameter $f_m$ was introduced: $K_m^1=K_m^2=K_m^3=ff_m$.
The earlier parameters $f$ and $h$ were set to 1.
Figs. 4 A and B show the speed gain of HSSAex2 relative to the GA and the absolute computing time of HSSAex2, respectively. Fig. 4 C shows
the average and variance of the mRNA copy number of the first gene in the first cell for the GA (blue) and HSSAex2 (red).

\section{Discussion and conclusion}

We have presented exact derivation and practical applications of a hybrid stochastic simulation algorithm (HSSA) that
can be, depending on system parameters, orders of magnitude faster than the Gillespie algorithm (GA), and highly accurate.
The principal behind the HSSA is the partitioning of a system of reactions into two groups, $A$ and $B$; the reactions in $A$
are simulated using a Gillespie-type algorithm, while the reactions in $B$ are described by the chemical master equation (CME).
We have derived exact formulas and equations which allow, in principal, any reaction network to be partitioned in
an arbitrary way, up to the condition that there exists
a subset of variables that is affected only by reactions in $A$. For biological systems such as gene regulatory networks (GRN),
this condition is nearly always satisfied. One way to violate it would be to have a fully connected reaction network in which every species of molecule
interacts with every other species of molecules -- which is rare at best. 

Although the prescribed steps of the HSSA are straight forward, carrying them
out may range from tractable, to difficult to impossible, depending on the specific partitioning of the system.
To demonstrate how a reaction network can be partitioned, and how the HSSA may be implemented in practice,
we chose a GRN comprised of three interacting genes and gave two detailed examples of how this particular system lends itself to
partitioning. In the first example, the reactions in $A$ were merely the transcription and degradation of the mRNA of all three genes (6 reactions in total);
while in the second example, we also included the reactions that change promoter states of all three genes (18 reactions in total). 
In carrying out the steps of the HSSA in both examples, we made the
assumption that the reactions consisting of translation, forward and backward homodimerization, and degradation of the monomers and homodimes for
all three genes, lead to fluctuations that are close to Poisson. This allowed us to set these fluctuations to zero, thereby reducing the complexity present in
the steps of the HSSA. Consequently, the information about the copy numbers for those species that were described via the CME was reduced to their averages.
To obtain information about the fluctuations, one can write down equations for the statistical moments for these species which would lead only to a marginal
loss of efficiency. This will be demonstrated in a future work.

The comparison in speed and accuracy presented in the ``Results" section establishes the HSSA as an extremely useful tool
for studying stochasticity in GRN, especially as implemented in example 2. Depending on the parameters, we found that the HSSAex2
was at least 11 times faster than the GA (Fig. 2, $h=10$, $f=1$) and at most 96 times faster (Fig. 2, $h=1$, $f=10$) for simulations of
a single cell. For an array of cells, the HSSAex2 was up to 445 times faster (Fig. 4 A, $f_m=5$, $\gamma_\lambda=10$, in the case of 4 cells).
The HSSAex1 did considerably worse compared to example 2. The best case scenario in single-cell simulations
was a speed gain factor of 14 (Fig. 2, $h=10$, $f=10$). Given the superior performance of
the implementation in example 2, we did not simulate a multi-cell array using HSSAex1.

It may seem counterintuitive that HSSAex1 be slower than HSSAex2, given that the former has fewer reactions to simulate compared
to the latter. However, when we consider the set of tasks that HSSAex1 has to perform in $B$, it becomes understandable. In particular, computing eigenvalues and
eigenvectors in Eq. (\ref{Qsol}) and searching for the solution to $Q(t)=\xi_1$ are computationally more expensive than evaluating the relatively
simple algebraic expressions in HSSAex2. However, as we saw for the last case in Fig. 2, the HSSAex1 performed slightly better than HSSAex2 due to the
faster promoter dynamics engendered by an increase in $h$. This emphasizes the dependence of the network topology and parameters on its partitioning.

Out of many types of oligomers, the system we chose to work with contained only homodimers. Addition of heterodimers and higher
ologomers is trivial in both HSSAex1 and HSSAex2: one only needs to modify Eqs. (\ref{Averagesns}) to include the formation, dissociation and
degradation of these species. Although these reactions would couple the hitherto separate equations for the proteins and homodimers belonging to different genes, the polynomial expansion
on $\tau$ would still be applicable. For this type of system, the HSSA would likely perform even better relative to the GA, given that the number of additional reactions such a coupling would introduce
is proportional to the square of the number of genes (for dimers and a fully connected protein interaction network). This would lead to a significant loss of efficiency for the GA, but
only a moderate one for the HSSA. 

In conclusion, we have derived exact expressions needed to partition an arbitrary reaction network for the
purpose of implementing an HSSA. We showed on a three-gene network how the system may be partitioned. The
two ways of partitioning lead to similar accuracy but significant overall difference in efficiency. The largest speed gain, compared to the GA,
reached a factor of 445 for an array of 4 identical cells. Given that the reactions of our system of choice are ubiquitous in systems biology,
we believe that the methodologies advanced in this paper will not only serve as preferred tools for discovering stochastic properties of large GRN,
but will also open doors to further research in the technical aspects of system partitioning.


\begin{thebibliography}{10}
\bibitem{Albert6} Work in progress (paper will follow in near future)
\end{thebibliography}
\end{document}